\documentclass[a4paper]{article}

\usepackage{INTERSPEECH2020}
\usepackage{enumitem, multirow,booktabs}
\usepackage{blindtext,letltxmacro,xcolor,xparse}
\usepackage{hyperref}
\usepackage{tablefootnote}

\newcommand\BUT{$^1$} 
\newcommand\TID{$^2$} 
\newcommand\UPF{$^{3}$} 
\newcommand\UDB{$^4$} 

\title{BCN2BRNO: ASR System Fusion for Albayzin 2020 Speech to Text Challenge}

\name{Martin Kocour\BUT, Guillermo C\'{a}mbara\TID$^{,}$\UPF, Jordi Luque\TID, David Bonet\TID, Mireia Farr\'{u}s\UDB, Martin~Karafi\'{a}t\BUT, Karel Vesel\'{y}\BUT and Jan ``Honza'' \v{C}ernock\'{y}\BUT}
\address{
\BUT Brno University of Technology, Speech@FIT, IT4I CoE\\
\TID Telef\'onica Research\\
\UPF Universitat Pompeu Fabra\\
\UDB Universitat de Barcelona
}
\email{ikocour@fit.vutbr.cz}

\begin{document}
\maketitle{}
\begin{abstract}

\noindent This paper describes joint effort of BUT and Telef\'onica Research on development of Automatic Speech Recognition systems for Albayzin 2020 Challenge.
We compare approaches based on either hybrid or end-to-end models.
In hybrid modelling, we explore the impact of SpecAugment\cite{maliddi:is:2016:specaug,park:IS:2019:specaug} layer on performance.
For end-to-end modelling, we used a convolutional neural network with gated linear units (GLUs).
The performance of such model is also evaluated with an additional n-gram language model to improve word error rates.
We further inspect source separation methods to extract speech from noisy environment (i.e. TV shows).
More precisely, we assess the effect of using a neural-based music separator named Demucs\cite{defossez2019demucs}.
A fusion of our best systems achieved 23.33\,\%~WER in official Albayzin 2020 evaluations.
Aside from techniques used in our final submitted systems, we also describe our efforts in retrieving high-quality transcripts for training.
\end{abstract}

\vspace{1em}

\section{Introduction}

Albayzin 2020 challenge is a continuation of the Albayzin 2018 challenges~\cite{lleida2019albayzin}, which has evaluations for the following tasks: Speech to Text, Speaker Diarization and Identity Asignement, Multimodal Diarization and Scene Description and Search on Speech. The target domain of the series is broadcast TV and radio content, with shows in a notable variety of Spanish accents.


This paper describes  BCN2BRNO's team Automatic Speech Recognition (ASR) system for IberSPEECH-RTVE 2020 Speech to Text Transcription Chal\-lenge, a joint collaboration between Speech@FIT research group, Telef\'onica Research (TID) and Universitat Pompeu Fabra (UPF). Our goal is to develop two distinct ASR systems, one based on a hybrid model \cite{TDNN-F}  and the other one on an end-to-end approach \cite{wav2letterWSJRecipe}, and complement each other through a joint fusion. 

We submitted one primary system and one contrastive system. The primary system\,--\,Fusion B\,--\,is a word-level ROVER fusion of hybrid ASR models and end-to-end models. It achieved 23.33\,\% WER on official evaluation dataset. However, the same result was accomplished by the contrastive system\,--\,Fusion A--, a fusion which comprises only hybrid ASR models. In this paper we describe both ASR systems, plus a post-evaluation analysis and experiments that lead to a better performance of the primary fusion. We also discuss the effect of speech enhancement techniques like background music removal or speech denoising.


\section{Data}

The Albayzin 2020 challenge comes with two databases: \emph{RTVE2018} and \emph{RTVE2020}. The RTVE2018 is the main source of training and development data, while the RTVE2020 database is used for the final evaluation of submitted systems.
RTVE2018 database~\cite{RTVE2018} comprises 15 different TV programs broadcast between 2015 and 2018 by the Spanish public television Radiotelevisión Española (RTVE). 
The programs contain a great variety of speech scenarios from read speech to spontaneous speech, live broadcast, political debates, etc. They cover also different Spanish accents, including Latin-American ones.
The database is partitioned into 4 different subsets: train, dev1, dev2 and test. 
The database consists of $569$ hours of audio data, from which $468$ hours are provided with subtitles (train set), and $109$ hours are human-revised (dev1, dev2 and test sets).
Both hybrid and end-to-end models utilize dev1 and train sets for training, while dev2 and test sets serve as validation datasets. RTVE2020 database~\cite{RTVE2020} consists of TV shows of different genres broadcast by the RTVE from 2018 to 2019. It includes more than $70$ hours of audio and it has been whole manually annotated. 

In addition, three Linguistic Data Consortium (LDC) corpora were used for training the language model in the hybrid ASR system:
\emph{Fisher Spanish Speech}, \emph{CALLHOME Spanish Speech} and \emph{Spanish Gigaword Third Edition}.

Fisher Spanish Speech~\cite{fisher_sp} corpus comprises spon\-ta\-ne\-ous telephone speech from $136$ native Caribbean Spanish and non-Caribbean Spanish speakers with full orthographic transcripts. 
The recordings consists of $819$ telephone conversations lasting up to $12$ minutes each. 

CALLHOME Spanish Speech~\cite{callhome_sp} corpus consists of $120$ telephone conversations between Spanish native speakers lasting less than 30 minutes. Spanish Gigaword Third Edition~\cite{giga_sp} is an extensive database of Spanish newswire text data acquired by the LDC. It includes reports, news, news briefs, etc. collected from 1994 through Dec 2010. We also downloaded the text data from Spanish Wikipedia.

The end-to-end model is trained on Fisher Spanish Speech, Mozilla's Common Voice Spanish corpus and Telef\'onica's Call Center in-house data (23 hours).
Mozilla's Common Voice Spanish~\cite{ardila2019common} corpus is an open-source dataset that consists of recordings from volunteer contributors pronouncing scripted sentences, recorded at 48kHz rate. The sentences come from original contributor donations and public domain movie scripts. The version of Common Voice corpus used for this work is 5.1, which has 521 hours of recorded speech. However, we have kept only speech validated by the contributors, an amount of 290 hours.



\subsection{Transcript retrieval}
The training data from RTVE2018 database includes many hours of subtitled speech. 
Although, the captions contain se\-ve\-ral errors. 
In the most cases captions are shifted by a few seconds, so a segment with correct transcript corresponds to a different portion of audio. 
This phenomenon also occurs in human-revised development and test sets. 
Another problem with subtitled speech is ``partly-said'' captions. 
This issue involves misspelled and unspoken words of the transcription.

Since the training procedure of the hybrid ASR is quite error-prone in case of misaligned labels, we decided to apply a transcript retrieval technique developed by Manohar, et al.~\cite{ManoharMGB2017}:
the closed-captions related to the same audio, i.e., the whole TV show, are first concatenated according to the original timeline. 
This creates a small text corpus containing a few hundreds of words.
The text corpus is used for training a biased $N$-gram language model (LM) with $N = 7$, so the model is biased only on the currently processed captions.
During decoding, the weight of the acoustic model (AM) is significantly smaller than the weight of LM, because we believe that the captions should occur in hypotheses. 
Then, the ``winning'' path is retrieved from the hypothesis lattice as the path that has a minimum edit cost w.r.t. the original transcript. 
Finally, the retrieved transcripts are segmented using the CTMs obtained from the oracle alignment (previous step). 
More details can be found in~\cite{kocourDIP,ManoharMGB2017}.

\begin{table}[htbp]
    \centering
    \caption{2-pass transcript retrieval.}
    \begin{tabular}{llllr}
        \toprule
        \textbf{Cleaning} & Train & Dev1 & Dev2 & Test\\
        \midrule
        Original            & $468$     & $60.6$    & $15.2$    & $36.8$ \\
        1-pass              & $99.4$    & $21$      & $7.5$     & - \\
        2-pass              & $234.2$   & $55.1$    & $14.3$    & $33.7$   \\ 
        \midrule
        \textbf{Recovered}  & $50\,\%$  & $91\,\%$  & $94\,\%$  & $92\,\%$ \\
        \bottomrule
    \end{tabular}
    \label{table:cleaning}
\end{table}

The transcript retrieval technique is applied twice. First, we train an initial ASR system on out-of-domain data, e.g., Fisher and CALLHOME. A system is used in the first pass of transcript retrieval. Then, a new system is trained from scratch on already cleaned data and the whole process of transcript retrieval is repeated again. Table~\ref{table:cleaning} shows how this 2-pass cleaning leads to recover almost all the manually annotated development data and half of the subtitled training data.

\begin{figure}[htbp]
    \centering
    \caption{Amount of cleaned audio per TV-show, in hours.}
    \includegraphics[width=0.45\textwidth]{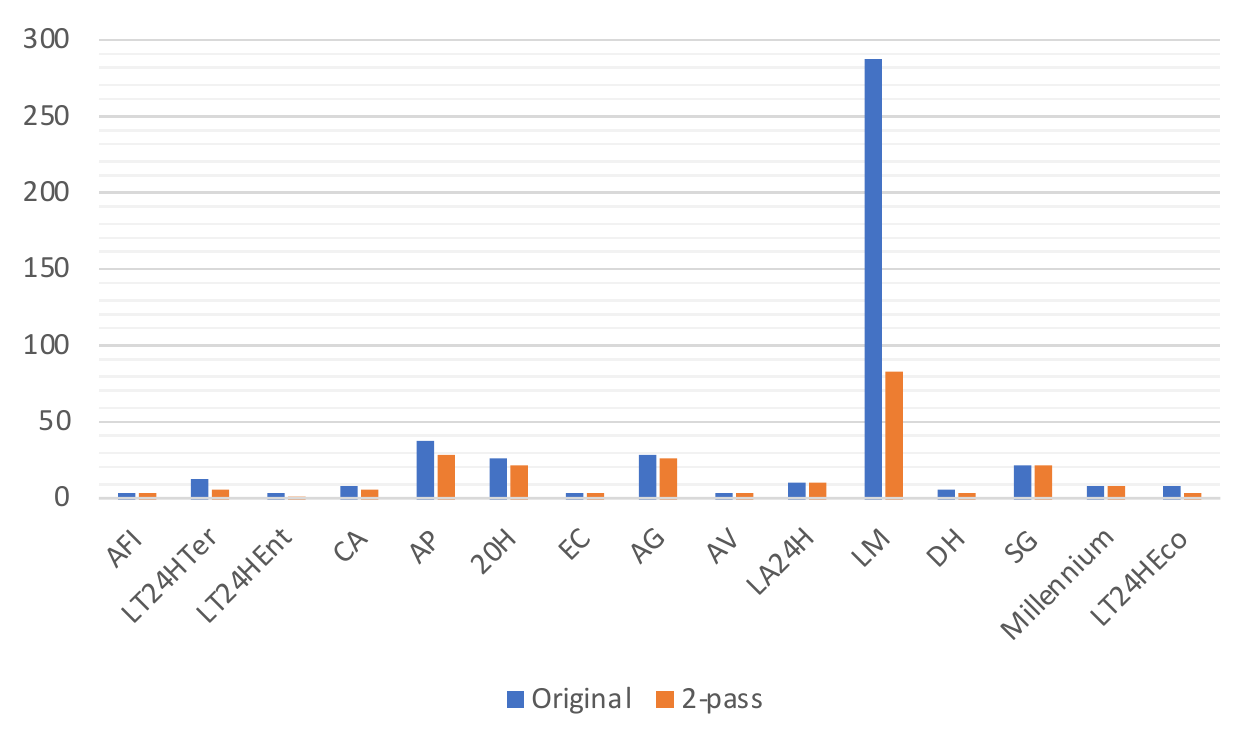}
    \label{fig:recovery}
\end{figure}

Figure~\ref{fig:recovery} depicts how many hours have been recovered in individual TV programs.
It also shows how data is distributed in the database.
Most speech comes from La-Mañana (LM) TV program.
We discarded most data in this TV program after 2-pass data cleaning.
It happened because this particular TV show was quite challenging for our ASR model.

\section{Hybrid speech recognition}

\subsection{Acoustic Model}

In all our experiments, the acoustic model was based on a hybrid Deep Neural Network\,--\,Hidden Markov Model architecture trained in Kaldi~\cite{kaldi}.
The NN part of the model contains 6~convolutional layers followed by 19~TDNN layers with semi-orthogonal factorization~\cite{TDNN-F}~(CNN-TDNNf). The input consists of 40-dim MFCCs concatenated with speaker dependent 100-dim i-vectors. 
Whole model is trained using LF-MMI objective function with bi-phone acoustic units as the targets.

In order to make our NN model training more robust, we introduced feature dropout layer into the architecture.
This prevents the model from overfitting on training data. In fact, it turned overfitting problem into underfitting problem. 
Thus, it leads to a slower convergence during training.
This is solved by increasing the number of epochs from 6 to 8 to balance the underfitting in our system.
This technique is also known as Spectral Augmentation. 
It was first suggested for multi-stream hybrid NN models in~\cite{maliddi:is:2016:specaug} and fully examined in~\cite{park:IS:2019:specaug}.

\subsection{Language Model}

We trained three different $3$-gram language models: Alb, Wiki and Giga. 
The names suggest which text corpus was used during training.
Albayzin LM was trained on dev1 and train sets from RTVE2018.
This text mixture contains $80$~thousand unique words in $0.5$~million sentences.
This small training text is not optimal to train $N$-gram LM, which is able to generalize well.
So we also included larger text corpora: Wikipedia and Spanish Gigaword.
These databases were further processed to get rid of unrelated text like advertisement, emoji, urls, etc.
This resulted into more than $2.5$~million fine sentences in Wikipedia and $20$~million sentences in Spanish Gigaword.
We experimented with $4$ combinations of interpolation:  Alb, Alb+Wiki, Alb+Giga, Alb+Wiki+Giga.

Our vocabulary consists of words from RTVE2018 database and from Santiago lexicon\footnote{\url{https://www.openslr.org/34/}}.
The pronunciation of Spanish words was extracted using public TTS model called E-speak~\cite{espeak}. 
The vocabulary was then extended by auxiliary labels for noise, music and overlapped speech.
The final lexicon contains around $110$~thousand words.

\subsection{Voice Activity Detection}

Voice activity detection (VAD) was applied on evaluation data in order to segment the audio into smaller chunks. 
VAD is based on feed-forward neural network with two outputs.
It expects 15-dimensional filterbank features with additional 3 Kaldi pitch features \cite{gharemani:ICASSP:2014:kaldi_pitch} as the input.
Features are normalized with cepstral mean normalization. 
More details can be found in~\cite{VeselyVad}.

\section{End-to-end speech recognition}

\subsection{Acoustic Model}

The end-to-end acoustic model is based on a convolutional architecture proposed by Collobert et al.~\cite{wav2letterWSJRecipe} that uses gated linear units (GLUs). Using GLUs in convolutional approaches helps avoiding vanishing gradients, by providing them linear paths while keeping high performances. Concretely, we have used the model from wav2letter's Wall Street Journal (WSJ) recipe. This model has approximately 17M parameters with dropout applied after each of its 17 layers. The WSJ dataset contains around 80 hours of audio recordings, which is smaller than the magnitude of our data ($\sim${600} hours). The LibriSpeech recipe ($\sim${1000} hours) provides a deeper ConvNet GLU based architecture, however we decided to use the WSJ one in order to reduce computational time and improve hyper-parameter fine-tuning of the network.


All data samples are resampled at 16kHz, and the system is trained with wav2letter++ framework. 
Mel-frequency spectral coefficients (MFSCs) are extracted from raw audio, using 80 filterbanks, and the system is trained using the Auto Segmentation criterion (ASG) \cite{wav2letterWSJRecipe} with batch size set to 4.
The learning rate starts at 5.6 and is decreased down to 0.4 after 30 epochs, where training is finished since no significant WER gains are achieved. 
From epochs 22 to 28 the system is trained also with the same train set, but adding the RTVE2018 train and dev1 samples with the background music cleaned by Demucs module \cite{defossez2019demucs}.
The last two epochs, from epoch 28 to epoch 30, are done incorporating further samples with background noise removed by Demucs and denoised by a neural denoiser \cite{defossez2020real}. 
This way, data augmentation with samples without background music and noise is done, to aid the network at training with samples with difficult acoustic conditions. Besides, the network is more likely to generalize audio artifacts caused by the denoiser and music separator networks, which is useful when using these to clean test audio. 


\subsection{Language Model}

Regarding the lexicon, we extract it from the train and validation transcripts, plus Sala lexicon \cite{moreno2002speechdat}. The resulting lexicon is a grapheme-based one with 271k 
words. We use the standard Spanish alphabet as tokens, plus the "ç" letter from Catalan and the vowels with diacritical marks, making a total of 37 tokens. 

The LM is a 5-gram model trained with KenLM \cite{kenLM} using only transcripts from the training sets:  RTVE2018 train and dev1, plus Common Voice, Fisher and Call Center. The resulting LM is described in this paper as \emph{Alb+Others}.

Fine-tuning of decoder hyperparameters is done via grid-search with RTVE2018 dev2 set. The best results are achieved with a LM weight of 2.25, a word score of 2.25 and a silence score of -0.35. This same configuration is then applied for evaluation datasets from RTVE2018 and RTVE2020.

\section{Experiments}

\subsection{Data cleaning}

Data cleaning by means of 2-pass transcript retrieval improves the performance of our models the most.
Table~\ref{table:cleaning} shows the effect of each pass. 
The $2$\textsuperscript{nd} pass helped to improve the accuracy by almost $2$\,\% in terms of WER.
We also ran the $3$\textsuperscript{rd} pass, but that did not help anymore. We simply did not retrieve more cleaned data from the original transcripts, just $3$ hours more. We could not train the models with the original subtitles, since these contained wrong timestamps.

\begin{table}[htbp]
    \centering
    \caption{Effect of 2-pass transcript cleaning evaluated on RTVE2018 test set.}
    \begin{tabular}{llcc}
        \toprule
        \multirow{2}{*}{AM} & \multirow{2}{*}{LM} & {Training} & WER [\%] \\
         & & data & Test\\
        \midrule
        \multirow{3}{*}{CNN-TDNNf} & \multirow{3}{*}{Alb} & 1-pass & 17.2\\
         & & 2-pass & 15.5\\
         & & 3-pass & 15.5\\
        \bottomrule
    \end{tabular}
    \label{table:results_cleaning}
\end{table}

\subsection{Speech Enhancement}\label{section:speech_enhancement} 
It is very common to find background music on TV programs, which can confuse our recognizer if it has a notorious presence. This brought us the idea of processing the audio through a Music Source Separator called Demucs~\cite{defossez2019demucs}. It separates the original audio into voice, bass, drums and others. By keeping only the voice component, we managed to significantly eliminate the background music, while maintaining relatively good quality in the original voice. 

We enhanced both validation sets in order to assess possible WER reductions. As seen in Table~\ref{tab:fusion}, this approach yielded a small increase in WER. We also tried applying a specialized denoiser \cite{defossez2020real} after background music removal, but the WER for dev2 increased in an absolute 1.6\%, compared to original system without enhancement. None of these two approaches (Demucs and Demucs+Denoiser) provided WER improvements at first, so we did not apply them for the end-to-end model used in the fusion. Although, the end-to-end, end-to-end + Demucs and end-to-end + Demucs + Denoiser models were submitted as separate systems by UPF-TID team, see Table~\ref{tab:final_results} for details.

\begin{figure}[ht]
    \centering
    \caption{Variation of the mean WER per TV show between using Demucs-cleaned or original samples on RTVE's 2018 test set. Negative values represent Demucs improvements. Note that only samples with SNR between -5 and 8 are enhanced.}
    \includegraphics[width=0.45\textwidth]{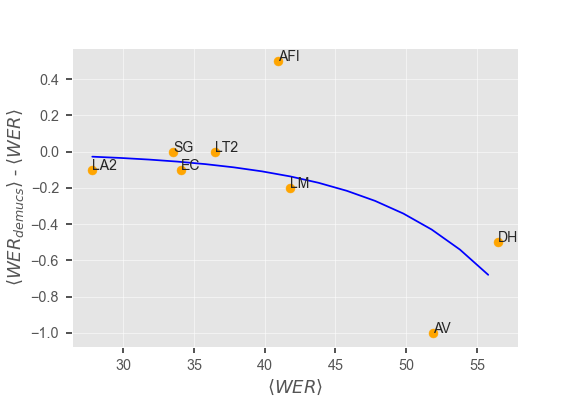}
    \label{fig:demucs}
\end{figure}

Our hypothesis is that not all the samples contain background music. Speech enhancement for already clean samples is detrimental because it causes slight degradation in the signal. Hence, we have evaluated the effects of applying music source separation to samples under certain SNR ranges, measured with the WADA-SNR algorithm ~\cite{kim2008robust}. The application of music separation on RTVE dataset is optimal for SNR ranges between -5 and 5 or 8 as it is shown in Table~\ref{table:demucs_snr}. Looking at Figure~\ref{fig:demucs}, best improvements are found at TV shows with higher WER (thus harder/noisier speech), e.g., AV, where most of the time speakers are in a car, or LM and DH, where music and speech often overlap. Other shows have slighter benefits, since these already contain good quality audio. The exception is AFI show, which is reported to have poor quality audio, so further audio degradation from Demucs might cause worse performance.


\begin{table}[htbp]
    \centering
    \caption{WER impact of cleaning speech signals between certain SNR ranges, using a music source separator. End-to-end ConvNet GLU model is used without LM, and percentage of cleaned samples are reported.}
    \begin{tabular}{lcccc}
        \toprule
         \multirow{1}{*}{SNR} & \multicolumn{2}{c}{Cleaned Samples [\%] } & 
         \multicolumn{2}{c}{Test WER [\%]} \\
     & 2018 & 2020 & 2018 & 2020 \\
     
        \midrule
        \multirow{1}{*}{$(-\infty, \infty)$} & \multirow{1}{*}{100} & 100 & 37.50 & 53.53\\
        \multirow{1}{*}{$(-\infty, 10)$} & 25.97 & 34.22 & -0.05 & -0.87\\
        \multirow{1}{*}{($-5$,  $10$)} & 25.84 & 31.33 & -0.05 & -0.88\\
        \multirow{1}{*}{($-5$,  $5$)} & 5.14 & 11.88 & -0.07 & \textbf{-1.03}\\
        \multirow{1}{*}{($-5$,  $8$)} & 14.95 & 22.11 & \textbf{-0.08} & -0.97\\
        \bottomrule
    \end{tabular}
    \label{table:demucs_snr}
\end{table}

\subsection{Spectral augmentation}

Table~\ref{tab:fusion} shows compared models with and without spectral augmentation.
The technique helps quite significantly. 
All models with feature dropout layer outperformed their counterparts with a quite constant $0.4\%$ absolute WER improvement on RTVE2018 test set and around $0.6\%$ on RTVE2018 dev2 set.

\subsection{Model fusion}\label{sec:fusion}

We also fuse the output of our best systems to further improve the performance. 
Overall results of our systems considered for the fusion are depicted in Table~\ref{tab:fusion}.
Since the models with spectral augmentation performed significantly better, we decided to fuse only these systems.
We analyzed two different approaches: a pure hybrid model fusion (Fusion A) and hybrid and end-to-end model fusion (Fusion B). 

Considering that the end-to-end model does not provide word-level timestamps, we had to force-align the transcripts with the hybrid ASR system in order to obtain CTM output.
The original 
word-level fusion was done using ROVER toolkit~\cite{rover}. Fusion B with end-to-end models performed slightly better than its counterpart Fusion A, despite the fact that the end-to-end models achieved worse results. This somehow proves the idea that the fusion can benefit from different modeling approaches.


\section{Final systems}

Table~\ref{tab:final_results} depicts the results on RTVE2020 test set.
For the end-to-end ConvNet GLU model, the performance drops around a 15\% WER when compared with previous results on development sets. Since the TV shows in such sets are also present in training dataset, our hypothesis is that the model slightly overfits to them. Therefore, when facing different acoustic conditions, voices, background noises and musics presented in RTVE2020 test set, the WER noticeable increases. Enhancing the test samples with Demucs or with Demucs+Denoiser yields a worse WER score, probably due to an inherent degradation of the signal.
A deeper analysis about more efficient ways to apply such enhancements has been done in section~\ref{section:speech_enhancement}.

Also, note that the submitted systems had a leak of dev2 stm transcripts in the LM, causing an hyperparameter overfitting during LM tuning. This caused a WER drop in all end-to-end systems, yielding WERs of 41.4\%, 42.3\% and 58.6\%. 
Table~\ref{tab:final_results} also displays the results of same systems with the leakage and LM tuning corrected in post-evaluation analysis.

\begin{table}[htb]
\centering
\caption{Overall results on RTVE2018 dataset with various language models and fusions.}
\label{tab:fusion}
\begin{tabular}{rllcc}
    \toprule
     & \multirow{2}{*}{AM} & \multirow{2}{*}{LM} & \multicolumn{2}{c}{WER [\%]} \\
     & & & Dev2 & Test\\
    \midrule
    1 & \multirow{4}{*}{CNN-TDNNf} & Alb & 14.1 & 15.5\\
    2 & & Alb + Wiki & 13.6 & 14.9\\
    3 & & Alb + Giga & 13.6 & 15.1\\
    4 & & Alb + Wiki + Giga & 13.5 & 15.0\\
    \midrule
    5 & \multirow{4}{*}{+ SpecAug} & Alb & 13.4 & 15.0\\
    6 & & Alb+Wiki & 12.9 & 14.5\\
    7 & & Alb+Giga & 13.0 & 14.7\\
    8 & & Alb+Wiki+Giga & 12.9 & 14.6\\
    \midrule
    9  & \multirow{2}{*}{ConvNet GLU} & None & 36.1 & 37.5\\
    10 & & Alb + Others & 20.8 & 20.7\\
     \midrule
    11 & \multirow{2}{*}{+ Demucs} & None & 36.4 & 37.5\\
    12 & & Alb + Others & 21.1 & 20.8\\
     \midrule
    13 & \multicolumn{2}{l}{Fusion A\qquad\emph{(row 5-8)}} & 12.9 & 13.7 \\
    14 & \multicolumn{2}{l}{Fusion B\qquad\emph{(row 5-8 and 10)}} & \textbf{12.8} & \textbf{13.3} \\
    \bottomrule
\end{tabular}
\end{table}

\begin{table}[htb]
\centering
\caption{Official and post-evaluation final results on RTVE2020 eval set for the submitted systems.}
\label{tab:final_results}
\begin{tabular}{llccc}
    \toprule
    \multirow{2}{*}{Model} &  \multicolumn{2}{c}{WER [\%]} \\
     & Official & Post-eval\\ 
    \midrule
    \multirow{1}{*}{CNN-TDNNf} & - & 24.3\\
    \multirow{1}{*}{+ SpecAug} & - & 23.5\\
    \midrule
    \multirow{1}{*}{ConvNet GLU } & 41.4\tablefootnote{Primary system of UPF-TID team.} & 36.2\\
    \multirow{1}{*}{+ Demucs } & 42.3\tablefootnote{First contrastive system of UPF-TID team.} & 37.9\\
    \multirow{1}{*}{+ Demucs + Denoiser } & 58.6\tablefootnote{Second contrastive system of UPF-TID team.} & 40.0\\
     \midrule
    \multirow{1}{*}{Fusion A} & 23.33 & 23.38\\
    \multirow{1}{*}{Fusion B} & \textbf{23.33} & \textbf{23.24}\\
    \bottomrule
\end{tabular}
\end{table}

\section{Conclusions}
In this paper we described two different ASR model architectures and their fusion.
We focused on improving the original subtitled data in order to train our models on high quality target labels.
We also improved the $N$-gram language model by incorporating publicly available text data from Wikipedia and Spanish Gigaword corpus from LDC. 
We have also successfully incorporated the spectral augmentation into our AM architecture.
Our best system achieved $13.3$\,\% and 23.24\,\% WER on RTVE2018 and RTVE2020 test sets respectively.

The performance of our hybrid system can be further improved by using lattice-fusion with Minimum Bayes Risk decoding\cite{swietojanski:ICASSP:2013:lattice_fusion}.
Another space for improvement is offered by adding a RNN-LM lattice-rescoring.
Our end-to-end model shows relatively competitive performance on RTVE2018 test set in comparison with its hybrid counterpart. However, its performance on RTVE2020 expose that the model was not able to generalize very well since this database turns out to contain slightly different acoustic conditions. Despite of this fact, the model still managed to improve the results in the final fusion with hybrid systems. An exploration on background music removal shows that it yields the best results for lower SNR ranges, thus having a different impact depending on the acoustic conditions of each TV show.


\bibliographystyle{IEEEtran}

\bibliography{mybib}

\end{document}